# Learning Efficient Representations for Keyword Spotting with Triplet Loss

Roman Vygon [1, 2][0000-0002-3684-7356] and Nikolay Mikhaylovskiy [1, 2][0000-0001-5660-0601]

[1] Higher IT School, Tomsk State University, Tomsk, Russia, 634050
[2] NTR Labs, Moscow, Russia, 129594
`{rvygon, nickm}@ntr.ai`

**Abstract.** In the past few years, triplet loss-based metric embeddings have become a de-facto standard for several important computer vision problems, most notably, person reidentification. On the other hand, in the area of speech recognition the metric embeddings generated by the triplet loss are rarely used even for classification problems. We fill this gap showing that a combination of two representation learning techniques: a triplet loss-based embedding and a variant of kNN for classification instead of cross-entropy loss significantly (by 26% to 38%) improves the classification accuracy for convolutional networks on a LibriSpeech-derived LibriWords datasets. To do so, we propose a novel phonetic similarity based triplet mining approach. We also improve the current best published SOTA for Google Speech Commands dataset V1 10+2 -class classification by about 34%, achieving 98.55% accuracy, V2 10+2-class classification by about 20%, achieving 98.37% accuracy, and V2 35-class classification by over 50%, achieving 97.0% accuracy.[1]

## 1 Introduction

The goal of keyword spotting is to detect a relatively small set of predefined keywords in a stream of user utterances, usually in the context of small-footprint device [1]. Keyword spotting (KWS for short) is a critical component for enabling speech-based user interactions for such devices [2]. It is also important from an engineering perspective for a wide range of applications [3]. In this article we show how the use of the triplet loss-based embeddings allows us to improve the classification accuracy of the existing small-footprint neural network architectures.

---

[1] Code is available at https://github.com/roman-vygon/triplet_loss_kws

### 1.1 Previous work on KWS

The first work on KWS was most likely published in 1967 [4]. Over years, a number of machine learning architectures for small-footprint KWS have been proposed (see, for example [5][6][7][8][9]. With the renaissance of neural networks, they become the architecture class of choice for KWS systems (see, for example, [1][2][10][11][12][13][14]). Probably, the only – but notable – exception from this trend is the very recent work of Lei et al. [15] that uses Tsetlin machines for keyword spotting for their extremely low power consumption.

Publication of the Google Speech Command dataset [16] have provided a common ground for KWS system evaluation and allowed for accelerating research. Further, we denote V1 and V2 versions 1 and 2 of this dataset, respectively. When publishing the dataset, Warden [16] have also provided a baseline model based on the convolutional architecture of Sainath and Parada [11], achieving the accuracy of 85.4% and 88.2% on V1 and V2, respectively. The related Kaggle competition winner has achieved 91% accuracy on V1.

Since the publication of the Google Speech Command dataset led to a vast corpus of work appearing in the past three years, we will only briefly discuss the most relevant recent work. Jansson [17] suggested an interesting fully-convolutional model working out of raw waveforms, but, probably, a bit ahead of time and did not improve on previous results. de Andrade et al. [3] have proposed an attention-based recurrent network architecture and achieved the SOTA on 2, 10, 20-word and full-scale versions of the dataset. Majumdar and Ginsburg [18] have published a lightweight separable convolution residual network architecture MatchboxNet, achieving the new SOTA of 97.48% on V1 and 97.63% on V2. Mordido et al. [19] have suggested an interesting improvement to MatchboxNet model, replacing 1x1-convolutions in 1D time-channel separable convolutions by constant, sparse random ternary matrices with weights in {-1; 0; +1}.

Rybakov, Kononenko et al. [20] tested many of the existing models and proposed a multihead attention-based recurrent neural network architecture, achieving a new SOTA of 98% on V2. Wei et al. [21] proposed a new architecture, EdgeCRNN, which is based on depthwise separable convolution and residual structure, apparently drawing inspiration from MatchboxNet [18] and Attention RNN [3], to achieve a slight improvement in accuracy and a SOTA of 98.05%. Tang et al. [22] have released Howl - a productionalized, open-source wakeword detection toolkit, explored a number of models and achieved nearly-SOTA accuracy.

### 1.2 Previous work on the use of triplet loss for the metric embedding learning

The goal of metric embedding learning is to learn a function $f: R^F \to R^D$, which maps semantically similar points from the data manifold in $R^F$ onto metrically close points in $R^D$, and semantically different points in $R^F$ onto metrically distant points in $R^D$ [23].

The triplet loss for this problem was most likely first introduced in [24] in the framework of image ranking:

$$l(p_i, p_i^+, p_i^-) = \{0, g + D(f(p_i), f(p_i^+)) - D(f(p_i), f(p_i^-))\} \tag{1}$$

where $p_i, p_i^+, p_i^-$ are the anchor image, positive image, and negative image, respectively, $g$ is a gap parameter that regularizes the gap between the distance of the two image pairs: $(p_i, p_i^+)$ and $(p_i, p_i^-)$, and $D$ is a distance function that can be, for example, Euclidean distance in the image embedding space:

$$D(f(P), f(Q)) = \|f(P) - f(Q)\|_2^2 \quad (2)$$

A similar loss function was earlier proposed by Chechik et al. in [25], but the real traction came to the triplet loss in the area of face re-identification after the works of Schroff et al. on FaceNet [26] and Hermans et al. [23].

In the speech domain, the use of triplet loss is more limited, but there still are several important works to mention. In particular, Huang J. et al [27], Ren et al. [28], Kumar et al. [29], and Harvill et al. [30] use triplet loss with varied neural network architectures for the task of the speech emotion recognition. Bredin [31] and Song et al. [32] use triplet-loss based learning approaches for the speaker diarization, and Zhang and Koshida [33] and Li et al. [34] – for the related task of speaker verification. Turpault et al. [35] propose a strategy for augmenting data with transformed samples, in line with more recent works in varied machine learning areas.

The most similar works to ours are probably [36], [37], [38], [39], and [40], but there are important differences with each of these works:

- Sacchi et al. [36] operate in the open-vocabulary setting, which required the authors to design a system with a common embedding for text and speech, while we concentrate on improving the quality of existing low-footprint architectures for closed-vocabulary keyword spotting
- Shor et al. [37] concentrate on building an unified embedding that works well for non-semantic tasks, while we concentrate on the semantic task of keyword spotting
- Yuan et al. [38] operate in a two-stage detection / classification framework and use a BLSTM network with a mix of triplet, reverse triplet and hinge loss
- Huh et al. [39] start from the same res15 model as we do, but primarily focus on detection metrics and use SVM for classification, so our classification metrics are significantly better
- Huang et. al [40] concentrate on Query-by-Example KWS application and adopt the softtriple loss - a combination of triplet loss and softmax loss

### 1.3 Our contributions

Our contributions in this work are the following:

- We show that combining two representation learning methods: triplet-loss based metric embeddings and a kNN classifier allows us to significantly improve the accuracy of CNN-based models that use cross-entropy to classify audio information and achieve the SOTA for the Google Speech Commands dataset
- We propose a novel batch sampling approach based on phonetic similarity that allows to improve F1 metric when classifying highly imbalanced datasets

## 2      Model Architectures

Most of the current state-of-the-art keyword spotting architectures are present in the work of Rybakov et al. [20], with the best model to date being the Bidirectional GRU-based Multihead Attention RNN. It takes a mel-scale spectrogram and convolves it with a set of 2D convolutions. Then two bidirectional GRU layers are used to capture two-way long term dependencies. The feature in the center of the bidirectional LSTM's output sequence is projected using a dense layer and is used as a query vector for the multi-head attention (4 heads) mechanism. Finally, the weighted (by attention score) average of the bidirectional GRU output is processed by a set of fully connected layers for classification.

We have mostly experimented with ResNet-based models res8 [22][1] and res15 [41][1]. The initial experiments have shown that RNN-based architectures show significantly worse results when trained for the triplet loss, so they were discarded in our later work. We used the encoder part of each of the models above to generate triplet-loss based embeddings, that are later classified using the K-Nearest Neighbor (kNN) algorithm.

### 2.1     Input Preprocessing

64-dimensional (for LibriWords) or 80-dimensional (for Google Speech Commands dataset) mel-spectrograms are constructed and stacked using a 25-millisecond window size and a 10-millisecond frame shift. Our implementation stacks all such windows within the one-second sample of Google Speech Commands. LibriWords samples are constrained to have a duration of 0.1-3 seconds.

### 2.2     Resnet architecture

Our resnet implementation is taken directly from [41] with very minor code changes and is depicted in **Fig. 1**. When working with triplet loss, the softmax layer is removed.

**Table 1.** Encoder model sizes for the key models studied.

|            | **Embedding dimension** | **Model encoder size, [K]** |
|------------|-------------------------|-----------------------------|
| Mh-Att-RNN | 256                     | 743                         |
| res8       | 128                     | 885                         |
| res15      | 45                      | 237                         |
| Att-RNN    | 128                     | 202                         |

**Table 1** above compares the model sizes for the main models studied.

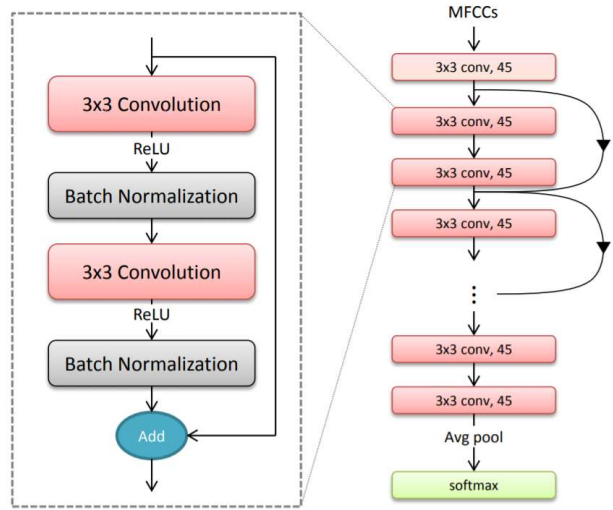

**Fig. 1.** res* architecture (from [1])

## 3 Experiments

### 3.1 Datasets and tasks

**SpeechCommands.** Google Speech Commands dataset Version 1 has 65K utterances from various speakers, each utterance 1 second long. Each of these utterances belongs to one of 30 classes corresponding to common words like "Go", "Stop", "Left", "Down", etc. Version 2 has 105K utterances, each 1 second long, belonging to one of 35 classes. The sampling rate of both datasets is 16kHz. In our experiments we have considered the following tasks based on these datasets [3][16]:

- Recognition of all 35 words using Google Speech Dataset V2
- Recognition of 10 words ("Yes", "No", "Up", "Down", "Left", "Right", "On", "Off", "Stop", and "Go") and additional labels for "Unknown" and "Silence" using either V1 or V2 datasets.

For these tasks and each architecture studied we have measured top-1 classification accuracy.

**LibriWords Datasets.** To further explore the possibilities of triplet loss models we needed a dataset that consists of a large number of different words to classify. Thus, we have used LibriSpeech [42] - a collection of 1,000 hours of read English speech. The dataset was split on the word level by Lugosch et al. [43]. Since LibriSpeech is aligned on sentence level only, the Montreal Forced Aligner [44] was used to obtain intervals for individual words. The alignments are available online [43]. Further we call this derived dataset LibriWords.

We have created four different versions of the dataset (LibriWords10, LibriWords100, LibriWords1000, LibriWords10000) that correspond to the first 10, 100 etc. words by popularity in the LibriSpeech 1000h corpus. For example, the LibriWords10 words are: "the", "and", "of", "to", "a", "in", "he", "I", "that", "was".

Durations of the words range from 0.03 seconds to 2.8 seconds, with mean duration of 0.28 seconds. The details on the datasets metrics are available in the Appendix 1. We have split the dataset into train\val\test in in 8:1:1 proportion, and tried to make sure this proportion holds for each word in the dataset. We release NeMo-like manifests for ease of use and reproduction. Since the motivation behind the dataset is to model real-life speech recognition scenarios, there was no further quality assurance on the data.

### 3.2 Approach to training models

**Batch sampling.** When working with Speech Commands and LibriWords10 datasets, to ensure a meaningful representation of the anchor-positive distances, following [26], we sample an equal number of objects from all the classes available. For unbalanced datasets with a large number of words, we also needed an efficient class-sampling method, otherwise the network will often train on irrelevant batches where embeddings of the words are already far from each other. To achieve better class selection we have used three sampling approaches:

- Uniform: sample *batch_size* classes randomly from a uniform distribution.
- Proportional: sample *batch_size* classes randomly from a distribution proportional to the word distribution in the dataset. Motivation behind this approach is twofold. First, the popular words are short (the, a, I)) so they are not easy to distinguish from the rest. Second, if you equally train on them, there will be the same amount of errors, and that's a lot in terms of the absolute value. (If we classify 2% of a popular word incorrectly, this would significantly spoil the metric for the entire dataset).
- Phonetic: Calculate a matrix of phonetic similarity for all the words in the dataset, sample batch_size/2 classes, then, for each sampled class add three random phonetically similar words (equally distributed) to the batch. Similarity score is calculated using SoundEx, Caverphone, Metaphone and NYSIIS algorithms [46].

**Comparing phonetic distance methods.** To compare the phonetic similarity algorithms, a model was trained using phonetic sampling only, while the similarity matrix was calculated with each of the methods separately. For each phonetic similarity algorithm, we have trained a model for two epochs on LibriWords10000 dataset. The results are listed in the **Table 4**. While the difference between the algorithms is not large, Metaphone leads in both the accuracy and F1.

Table 2. Phonetic similarity metric comparison

| Method | Accuracy | F1 |
|---|---|---|
| CaverPhone | 64.0 | 41.0 |
| NYSIIS | 63.8 | 40.5 |
| Soundex | 64.8 | 43.2 |
| Metaphone | 65.3 | 43.8 |

**Phonetic distance.** On LibriWords dataset, we used a weighted average of distances calculated using all 4 algorithms weighted as follows:

$$D_{Phonetic} = D_{Soundex} * 0.2 + D_{caverphon} * 0.2 + D_{metaphone} * 0.5 + D_{nysiis} * 0.1$$

The weights reasonably reflect the efficiency of each method as per **Table 2**. The optimal use of these algorithms is a matter of future research, for example, we had to adjust manually the distances of a handful of pairs of words: e.g. the pair "know-no" had a large distance while being similar. The problem was found while analyzing the confusion matrix.

We have evaluated these three triplet mining approaches alone and in combinations, mixing them with equal probabilities. Thus, for example, Uniform+Phounetic in the **Table 3** below means 50% probability to sample with the Uniform approach, and 50% with the Phonetic approach, and Uniform+Proportional +Phonetic means 1/3 probability to sample with the Uniform approach, 1/3 with the Phonetic approach, and 1/3 with the Proportional approach.

The results in the **Table 3** show that the proportional sampling method improves the accuracy by increasing the score of more popular words while the phonetic sampling method improves the F1 metric due to better classification of difficult pairs like "at"-"ate", "an"-"anne". Uniform sampling usage is essential as one of the sampling strategies, as it provides the proper class coverage.

Table 3. The effects of the different sampling strategies for triplet loss of res15 model on LibriWords10000

| Method(s) | Accuracy | F1 |
|---|---|---|
| Uniform | 79.4 | 0.72 |
| Proportional | 77.1 | 0.61 |
| Phonetic | 76.9 | 0.73 |
| Uniform+Phonetic | 78.9 | **0.76** |
| Uniform+Proportional | **81.2** | 0.74 |
| Proportional+Phonetic | 80.0 | 0.72 |
| Uniform+Proportional +Phonetic | 80.8 | 0.75 |

**Triplet selection.** An important part of TL models is the selection of triplets used to calculate the loss, since taking all possible triplets from a batch is computationally expensive. We have used a randomized approach to the online batch triplet mining based on [23], where the negative sample to a hard pair of the anchor and a positive sample is selected randomly from the set of negative samples resulting in non-zero loss. Our initial experiments have shown that this modification of the online batch triplet mining performs better than hard or semi-hard batch loss options.

**Optimization and training process.** Baseline models were trained until they reached a plateau on a validation set. We monitored the validation accuracy of triplet loss models each 1k batches and stopped the training process if the accuracy didn't increase for more than .1% for 3 consecutive times. The number of epochs is listed in the **Table 4** below.

Table 4. The number of epochs models were trained for

|  | TL, epochs | Baseline, epochs |
|---|---|---|
| Speech Commands | 30 | 30 |
| Libri10 | 10 | 30 |
| Libri100 | 5 | 10 |
| Libri1000 | 5 | 7 |
| Libri10000 | 3 | 5 |

Three augmentation techniques were used:

1. Shifting samples in range (-100ms; +100ms).
2. SpecAugment.
3. Adding background noise from audio files in Google Speech Commands Dataset.

The decrease in epochs for larger datasets is due to class-imbalance – triplet models sample classes directly, so instead of seeing all objects in the dataset it sees the same number of objects, but distributed more evenly between classes. The baseline, cross-validation based models converge to predict the most popular words well, while ignoring the rest. One can see this from the low F1 metric on LibriWords10000 dataset. The batch size was 35*10 for TL-res8, 35*4 for TL-res15 and 128 for the baseline models. Training was done using the Novograd [47] algorithm with initial learning rate of 0.001 and cosine annealing decay to 1e-4.

**Influence of kNN.** We have tested kNN for several values of k, and have found that for LibriWords the best performing value varies depending on the dataset size, while for Speech Commands the best performing value was $k=5$ (see **Table 5**).

As the model size is of a great concern for the keyword spotting application, and for the larger datasets kNN part of the model can take a lot of memory, we have also studied the effect of kNN quantization available from [12] on the size, speed and accuracy of the resulting model, varying the number of segments for the Product Quantizer.

Table 5. Classification accuracy for res15 model triplet loss embeddings with kNN classification for various k

| k | 1 | 5 | 10 | 30 |
|---|---|---|---|---|
| Speech Commands V2 / 12 | 98.18 | **98.37** | 98.27 | 98.29 |
| LW10 | 89.91 | 91.48 | **91.74** | 91.72 |
| LW100 | 83.93 | 86.53 | 86.9 | **86.98** |
| LW1000 | 80.43 | 83.82 | 84.29 | **84.37** |
| LW10000 | 77.57 | 80.82 | **81.17** | 80.62 |

For each dataset/task there is an optimal number of segments that reduces accuracy by 1.6% - 13.6%, and reduces the memory consumption by a factor of 7 to 13.

We should note that the use of kNN is essential for the accuracy we achieve. We have tried to replace kNN with a two-layer fully-connected network with ReLU between the layers, but the results were drastically worse. Specifically, we experimented with intermediary dimensions of 64, 128 and 256 between the two fully connected layers (see **Table 6**). We have frozen the same triplet loss based encoder as used with kNN and optimized each fully-connected decoder with a cross-entropy loss using Novograd optimizer for 30 epochs with cosine annealing from 1e-3 to 8e-5. The resulting accuracy was around 90% and F1 around 82% independently of the intermediary dimension. This means that the embeddings generated by the triplet loss are not linearly separable and using kNN is really critical for high-quality decoding the triplet loss embeddings.

**Table 6.** Classification accuracy and F1 for res15 model triplet loss embeddings with kNN and fully-connected network classifiers

| Metric | kNN | FC64 | FC128 | FC256 |
|---|---|---|---|---|
| Accuracy | 98.37 | 90.04 | 89.88 | 90.16 |
| F1 | 0.98 | 0.822 | 0.821 | 0.825 |

## 4 Results and Discussion

The results below were obtained by training a model for 3 different runs in each scenario and averaging the results to avoid the "lucky seed" effect. We can see that triplet loss + kNN based models provide better accuracy than baseline ones, achieve state of the art results on Speech Commands dataset, while being more lightweight and faster in convergence than the mh-att-rnn [20] model.

In particular, triplet loss + kNN based models improve the accuracy on the datasets studied by 25% to 38% and F1 measure by 16% to 57% compared to extremely strong crossentropy based baselines (see **Table 7**). The bigger the number of classes in the dataset, the bigger the difference between crossentropy and triplet loss based classifiers. Our res15 network trained with triplet loss and kNN classifier achieves state of the art on Google Speech Commands datasets V2/35, V2/12 and V1/12, improving the best previously published results [3][21][22] by 50%, 16% and 34% respectively (**Table 8**).

**Table 7.** Comparison of accuracy and F1 measure of triplet loss and crossentropy loss based res15 models

| Task | Triplet Loss | | Crossentropy | | Relative improvement | |
|---|---|---|---|---|---|---|
| | *Accuracy, %* | *F1* | *Accuracy, %* | *F1* | *Accuracy, %* | *F1,%* |
| Speech Commands V2 35 | 97.0 | 0.965 | 95.96 | 0.955 | 25.74 | 22.22 |
| Speech Commands V2 12 | 98.37 | 0.980 | 97.8 | 0.963 | 25.91 | 45.95 |
| Speech Commands V1 12 | 98.56 | 0.978 | 97.7 | 0.967 | 37.39 | 33.33 |
| LibriWords10 | 91.7 | 0.90 | 88.8 | 0.88 | 26.25 | 16.67 |
| LibriWords100 | 86.9 | 0.87 | 82.3 | 0.81 | 25.99 | 31.58 |
| LibriWords 1000 | 84.3 | 0.86 | 78.2 | 0.78 | 27.94 | 36.36 |
| LibriWords 10000 | 81.2 | 0.75 | 69.3 | 0.41 | 38.66 | 57.63 |

Table 8. Model accuracy comparison on Google Speech Commands dataset tasks

| Model | Loss | Model Size, KB | V2 35 accuracy | V2 12 accuracy | V1 12 accuracy |
|---|---|---|---|---|---|
| res8 (ours) | Triplet | 901 | 95.33 | 97.48 | |
| res8 (ours) | Crossentropy | 885 | 95.25 | 97.39 | 96.03 |
| res15 (ours) | Triplet | 252 | **97.0** | **98.37** | **98.56** |
| res15 (ours) | Crossentropy | 237 | 95.96 | 97.8 | 97.7 |
| EdgeCRNN [21] | Crossentropy | | | *98.05* | |
| Mh-Att-RNN [20] | Crossentropy | 743 | | 98.0 | |
| Attention RNN [3] | Crossentropy | 202 | *93.9* | | |
| res8 (Howl) [22] | Crossentropy | | | | *97.8* |

# 5    Acknowledgments

The authors are grateful to

- colleagues at NTR Labs Machine Learning Research group for the discussions and support;
- Prof. Sergey Orlov and Prof. Oleg Zmeev for the computing facilities provided;
- Nikolay Shmyrev for pointing out to the works [38], [39].